\documentclass[12pt,preprint]{aastex}
\shorttitle{Atmospheric Parameters of Metal-Rich Stars}
\shortauthors{Robinson et al.}
\begin{document}

\title{The N2K Consortium. VII. Atmospheric Parameters of 1907
Metal-Rich Stars: Finding Planet-Search Targets}

\author{Sarah E. Robinson\altaffilmark{1},
S. Mark Ammons\altaffilmark{1},
Katherine A. Kretke\altaffilmark{1},
Jay Strader\altaffilmark{1},
Jeremy G. Wertheimer\altaffilmark{1},
Debra A. Fischer\altaffilmark{2}, and
Gregory Laughlin\altaffilmark{1}}

\altaffiltext{1}{University of California Observatories/Lick
Observatory, Department of Astronomy and Astrophysics, University of
California at Santa Cruz, Interdisciplinary Sciences Building, Santa
Cruz, CA 95064, ser@ucolick.org, ammons@ucolick.org,
kretke@astro.ucsc.edu, jwerthei@ucolick.org, laughlin@ucolick.org}

\altaffiltext{2}{Department of Physics \& Astronomy, San Francisco State
University, San Francisco, CA 94132, fischer@stars.sfsu.edu}

\begin{abstract}
We report high-precision atmospheric parameters for 1907 stars in the
N2K low-resolution spectroscopic survey, designed to identify metal-rich
FGK dwarfs likely to harbor detectable planets.  284 of these stars are
in the ideal temperature range for planet searches, $T_{\rm eff} \leq
6000$K, and have a $10\%$ or greater probability of hosting planets
based on their metallicities.  The stars in the low-resolution
spectroscopic survey should eventually yield $> 60$ new planets,
including 8-9 hot Jupiters.  Short-period planets have already
been discovered orbiting the survey targets HIP 14810 and HD 149143.
\end{abstract}


\section{Introduction}
\label{introduction}

The peak year for planet discovery by radial velocity searches was 2002,
with 34 new planets discovered.  Since then, the planet discovery rate
has flattened out, with 27 new planets discovered in 2004 and another 27
in 2005\footnote{Source: Interactive Extra-solar Planets Catalog,
http://vo.obspm.fr/exoplanetes/encyclo/catalog.php}.  The volume within
25 pc of the Sun---$V < 7$ for a Solar-type star---has been thoroughly
searched for short-period giant planets, or hot Jupiters.  Future planet
searches focusing on this volume of space will be aimed at either
low-mass planets, as in the forthcoming Automated Planet Finder survey
at Lick Observatory, or long-period planets, as in the \cite{nakajima05}
coronagraphic-adaptive optics search for brown dwarfs and planets around
nearby ($d < 20$ pc) young stars.  Searches for new short-period planets
must push out to larger distances and fainter stars in order to add to
the 48 planets with periods $P < 7$ d (the upper limit for a tidally
circularized orbit) known at the time of this writing.

A primary reason the N2K Consortium focuses on discovering hot Jupiters
is their high probability of performing detectable transits.  Since the
radii of transiting planets can be measured directly, they provide
valuable information about planetary composition.  Indeed, planetary
interior models indicate that HD 149026 b, while having an observed mass
comparable to that of Saturn, has the largest solid core, $70
M_{\oplus}$, of any known planet (Sato et al. 2005; see also Fortney et
al. 2005).  In comparison, Jupiter has a core mass of $0 - 11
M_{\oplus}$ and Saturn has a core mass of $9 - 22 M_{\oplus}$ (Saumon \&
Guillot 2004).  It is also important to quickly locate short-period
planets orbiting bright parent stars because these objects can be
profitably observed with the {\it Spitzer Space Telescope} during its
limited cryogenic lifetime.  \cite{williams06} show that {\it Spitzer}
Infrared Array Camera secondary eclipse light curves can probe the
variation of hot Jupiter thermal emission across the surface of the
planet---for example, day-night temperature difference and presence of
hot or cold spots.

The N2K project (Fischer et al. 2005) was created to facilitate the
detection of new exoplanets, especially hot Jupiters, by identifying the
``Next Two Thousand'' metal-rich stars suitable for precise
radial-velocity measurements.  The observational efficiency of the
first-generation California-Carnegie planet search (Marcy et al.  2005)
of $V < 7$ stars was limited by readout and telescope slew time---total
$\sim 1$ minute per target with the HIRES spectrograph.  However, for
stars with $V > 7$, exposure time becomes the limiting factor on
observing efficiency.  To keep the planet detection rate per hour of
scientifically valuable short-period planets on par with its 2002 peak,
we must make sure each target star has a high probability of harboring a
detectable planet.  The first generation of planet searches uncovered
the planet-metallicity correlation (Gonzalez 1997): doubling the metal
content of a Solar-type star leads to a fourfold increase in its
probability of harboring a detectable planet (Fischer \& Valenti 2005).
By measuring the metallicity of thousands of FGK dwarfs in the Solar
neighborhood with published photometry and small-telescope observations,
we can identify the most productive targets for the ongoing
Keck/Subaru/Magellan (hereafter KSM) planet search.

The N2K consortium sieves targets for the KSM planet search using
successive metallicity estimates of increasing precision.  The first
step in the N2K pipeline is identifying FGK dwarfs in the Hipparcos
catalog (Perryman et al. 1997) that have not already been searched for
planets and have either 2MASS $JHK$ photometry (Skrutskie et al. 2006)
or $ubvy\beta$ photometry (N\"{o}rdstrom et al. 2004).  As most
Hipparcos stars have $V < 10$ and stars with $V < 7$ have, with few
exceptions, already been searched for planets, most of our candidate
stars have apparent magnitudes $7 < V < 10$.  For stars with Hipparcos
$BV$ and 2MASS $JHK$ photometry, we use the empirical relation between
broadband colors and metallicity calculated by \cite{mark} to estimate
[Fe/H].  This calibration has precision $ \sigma = 0.17$ dex for stars
with $V < 9$.  For the stars with $ubvy\beta$ photometry, we use the
[Fe/H] estimates reported in the \cite{nordstrom} compilation, most of
which were calculated using the \cite{sn89} $ubvy$-metallicity
calibration, with precision $\sigma = 0.13$ dex.

Stars with metallicity estimates of [Fe/H] = 0.0 dex or higher from
either broadband or $ubvy\beta$ photometry proceed to the second level
of N2K screening.  At this level, we obtain a low-resolution optical
spectrum of each star and measure the Lick indices, which are broad
atomic and molecular features between 4000 and 6000 \AA.  We then use
the empirical calibrations reported by Robinson et al. (2006, hereafter
Paper 1) that give [Fe/H], $T_{\rm eff}$ and log $g$ as functions of
selected Lick indices (Worthey et al. 1994).  The relatively high
precision of these calibrations---$\sigma_{\rm [Fe/H]} = 0.07$ dex,
$\sigma_{T_{\rm eff}} = 82$K, and $\sigma_{\log \, g} = 0.13$
dex---enables us to create clean target lists composed of metal-rich,
cool stars with high probabilities of planet detection for the ongoing
Keck/Subaru/Magellan planet search.  Note that, as our low-resolution
spectra were obtained at Kitt Peak National Observatory, there is a
large, unsurveyed population of Hipparcos/2MASS stars with declinations
below $-20^o$.  The Keck planet search also includes bright stars for
which the photometric metallicity estimates were precise enough to skip
low-resolution screening.

The first planet discovered by the N2K Consortium was HD 88133 b
(Fischer et al. 2005), a Saturn-mass planet with $P = 3.41$ d.  The next
discovery was the transiting hot Saturn orbiting HD 149026 (Sato et al.
2005).  The N2K consortium then reported two more short-period planets,
HD 149143 b and HD 109749 b (Fischer et al.  2006).  HD 149143 b is a
hot Jupiter, with minimum mass $M \sin i = 1.33 M_J$ and $P = 4.072 \:
{\rm d}$.  Finally, \cite{wright07} discovered two planets orbiting the
N2K target HIP 14810, the hot-Jupiter b component ($P = 6.67 \: {\rm
d}$) and the long-period c component.  In this paper, we report the
results of our low-resolution spectroscopic survey of 1907 stars.  Two
of N2K's newly discovered planet hosts, HD 149143 and HIP 14810, were
part of the low-resolution spectroscopic survey.


\section{Observations and Lick Index Measurements}
\label{obs}

Our observations were taken at the 2.1m telescope at Kitt Peak National
Observatory during three observing runs, UT dates 2004 August
27-September 2, 2005 March 26-April 1, and 2005 April 23-29.  A fourth
observing run, 2005 February 10-16, was rained out.  We used the GoldCam
spectrograph with a 600 lines mm$^{-1}$ grism blazed at 4900 \AA.  The
spectral coverage was 3800-6200 \AA\ with $R = 1360$ (FWHM = 3.7 \AA) at
5000 \AA.  A typical spectrum has S/N $\sim 230$ per resolution element
(120 per \AA) in the Ca4227 line, the shortest-wavelength index
measured, increasing to S/N $\sim 380$ per resolution element (200 per
\AA) in the Na D index.  As Lick indices are independent of absolute
flux levels (Worthey \& Ottaviani 1997), our spectra were not flux
calibrated.

Since our observing program was designed to survey as many potential
planet-search targets as possible, we did not take a comparison-lamp
spectrum at each telescope position.  Rather, we obtained wavelength
solutions accurate to $\sigma \sim 4$ \AA\ by observing comparison lamps
only at the beginning, middle and end of each night.  Following the
method of Paper 1, we used an unsharp masking algorithm to find the
center of each spectral line used in the Lick indices-atmospheric
parameter calibrations.  We smoothed each spectrum using a Gaussian
low-pass filter and subtracted the smoothed spectrum from the original
spectrum.  We then searched the unsharp-masked spectrum for local minima
with 12 \AA\ of each known line center.  Comparing line centers found by
the automatic recentering program with those measured by hand using
Gaussian-fit tools in IRAF for three spectra led us to estimate an error
of $\pm 2$ \AA\ in our recentered wavelength solutions.  According to
\cite{W94}, the contribution of wavelength errors of this magnitude to
errors in Lick indices is negligible.

The calibrations reported in Paper 1 use the bandpass definitions of
\cite{trager98} for the indices Ca4227, G4300, Fe4383, Fe4531, Fe4668,
H$\beta$, Fe5015, Mg$_2$, Mg {\it b}, Fe5270 and Na D; and the bandpass
definition of \cite{wo97} for H$\gamma_F$.  We measured Lick indices in
our spectra using the publicly available {\tt indexf}
code,\footnote{Created by Cardiel, Gorgas, \& Cenarro, released on 2002
July 11} which incorporates the error analysis techniques of
\cite{cardiel98}.  Since the Paper 1 calibrations are based in part on
observations with slightly lower resolution than the original IDS
spectra, we did not smooth our spectra to match the IDS resolution.
Measuring Lick indices using spectra with lower resolution than the
original IDS spectra slightly increases the random error in each index,
(Worthey \& Ottaviani 1997), but does not add any systematic errors.

\begin{figure}
\plotone{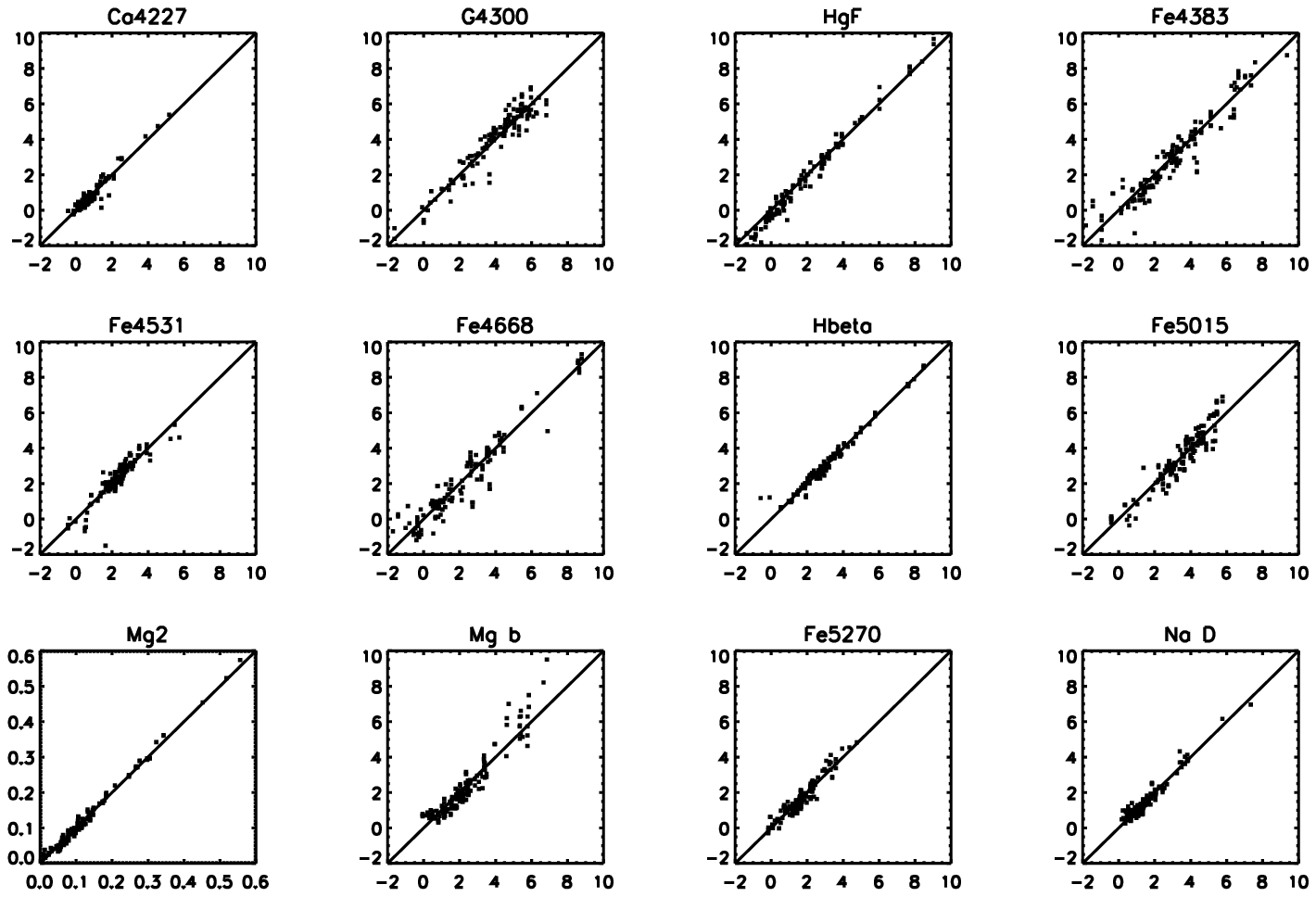}
\caption{Comparison between observed equivalent widths (y-axis) and
those published in \cite{W94} and \cite{wo97} (x-axis) for the 12
indices used in the Paper 1 fits to stellar atmospheric parameters.
Although separate transformations to the Lick system were calculated for
each observing run, our program has nearly uniform precision in
measuring Lick indices.  All Lick standards observed in our program are
shown together.}
\label{lickmatch}
\end{figure}

We transformed our data to the Lick system using observations of Lick
standard stars, which have indices reported in \cite{W94} and
\cite{wo97}.  79 observations of 62 standard stars were obtained during
the observing run in August 2004; 24 observations of 23 stars were
obtained during the March 2005 run; and 38 observations of 27 stars were
obtained during the April 2005 run.  The observed Lick standards were
mainly FGK dwarfs, matching the spectral types of our program stars, but
a few B and A-type dwarfs were also observed in parts of the sky where
FGK Lick standards were not available.  For each index, we used
least-squares analysis to find a linear fit between the published
equivalent width and the equivalent width measured from our data.  In
order for the fits to metal lines not to be biased by extremely
metal-poor stars, data points that were more than 3 standard deviations
away from the line of best fit were rejected and the fits were computed
again.  Rejecting deviant points also kept cool stars with no
discernible Balmer absorption from  biasing the fits to the indices
measuring Balmer lines, H$\gamma_F$ and H$\beta$.  Since the alignment
of the GoldCam spectrograph changes slightly each time it is taken down
and re-mounted on the telescope, we computed separate transformations to
the Lick system for each observing run.  The index measurement errors
and transformations from observed to published Lick indices are given in
Table \ref{licktrans}, and the Lick indices for our survey targets are
given in Table 2.  Figure \ref{lickmatch} compares our measured
Lick indices with published values for all the Lick standard stars in
our sample.

Rapidly changing temperatures on 2005 April 27 led to an unstable
telescope focus, and suboptimal spatial profiles of spectra obtained
that night.  Although spectra with wide spatial profiles as a
consequence of changing focus have reduced signal-to-noise ratio in each
line, the data obtained this night are still above the minimum S/N = 100
per \AA.  We see no systematic offsets in Lick index measurements for
the standard stars observed that night, and conclude that the accuracy
of data from 2005 April 27 is unimpaired.

\section{Measuring Atmospheric Parameters}
\label{measurement}

\begin{figure}
\plotone{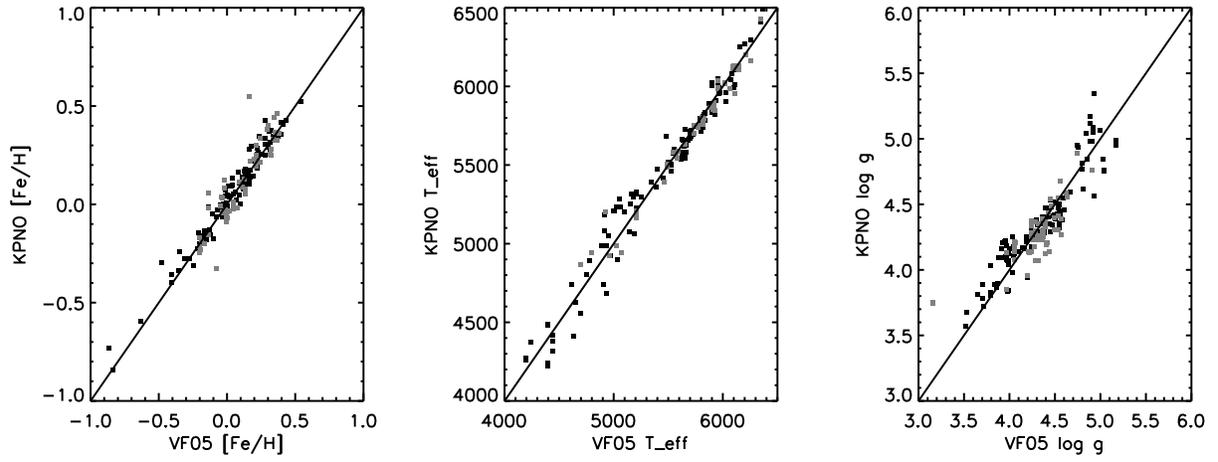}
\caption{Test of Paper 1 calibrations.  Stars that were used to build
the calibrations are shown in black, and stars that were reserved for
testing the calibrations are plotted in gray.  The solid line shows a
perfect 1:1 correspondence between our calibrations and VF05
measurements.}
\label{caltestscatter}
\end{figure}

During our KPNO observing program, we surveyed and measured the
atmospheric parameters of 1907 FGK dwarfs identified as metal-rich by
either \cite{mark} or \cite{nordstrom}.  [Fe/H], $T_{\rm eff}$ and log
$g$ were measured using the calibrations presented in Paper 1, which
were built by obtaining low-resolution spectra of stars in the Valenti
\& Fischer (2005, hereafter VF05) planet-search catalog and finding
empirical relations between selected Lick indices and the atmospheric
parameters reported in VF05.  In the Paper 1 fits, $T_{\rm eff}$ is
given by a linear combination of Lick indices; [Fe/H] is a linear
combination of Lick indices and $T_{\rm eff}$; and log $g$ is given by
linear terms in each of the Lick indices and $T_{\rm eff}$ plus one
nonlinear term, $T_{\rm eff} ({\rm H}\gamma_F + {\rm H}\beta)$.  To
verify the precision and accuracy of the fits in Paper 1, we obtained
191 observations of 127 stars in the VF05 catalog during the 3 KPNO
observing runs.  By comparing the atmospheric parameters measured from
KPNO spectra with the VF05 values, we could compare the true performance
of the [Fe/H], $T_{\rm eff}$ and log $g$ calibrations with the published
uncertainties.  Figure \ref{caltestscatter} gives scatter plots showing
the performance of each calibration.  Stars that were included in the
training set used to build the Paper 1 calibrations are shown in black,
and stars that were used only for testing the calibrations (``test
set'') are shown in gray.  A visual inspection of Figure
\ref{caltestscatter} reveals that the calibrations accurately reproduce
the VF05 atmospheric parameters.

In Paper 1, the calibration errors are modeled by fitting a Gaussian to
the residuals $({\rm [Fe/H]}_{\rm KPNO}) - ({\rm [Fe/H]}_{\rm VF05})$.
According to the Gaussian error model, $68\%$ or more of the stars in
any test set should have atmospheric parameter estimates within $1
\sigma$ of the VF05 values if the calibrations are performing within the
published error estimates.  The test set in this work consists of 79
observations of 48 stars, and the Paper 1 calibration uncertainties are
$\sigma_{T_{\rm eff}} = 82$K, $\sigma_{\rm [Fe/H]} = 0.07$ dex, and
$\sigma_{\log \, g} = 0.13$ dex.  $66\%$ of the log $g$ measurements are
within $1 \sigma$ of the VF05 values; $72\%$ of [Fe/H] measurements are
less than $1 \sigma$ from the VF05 values; and fully $91\%$ of $T_{\rm
eff}$ measurements are within $1 \sigma$ of VF05 values, indicating a
possible slight overestimation in our reported error on $\sigma_{T_{\rm
eff}}$.

\begin{figure}
\plotone{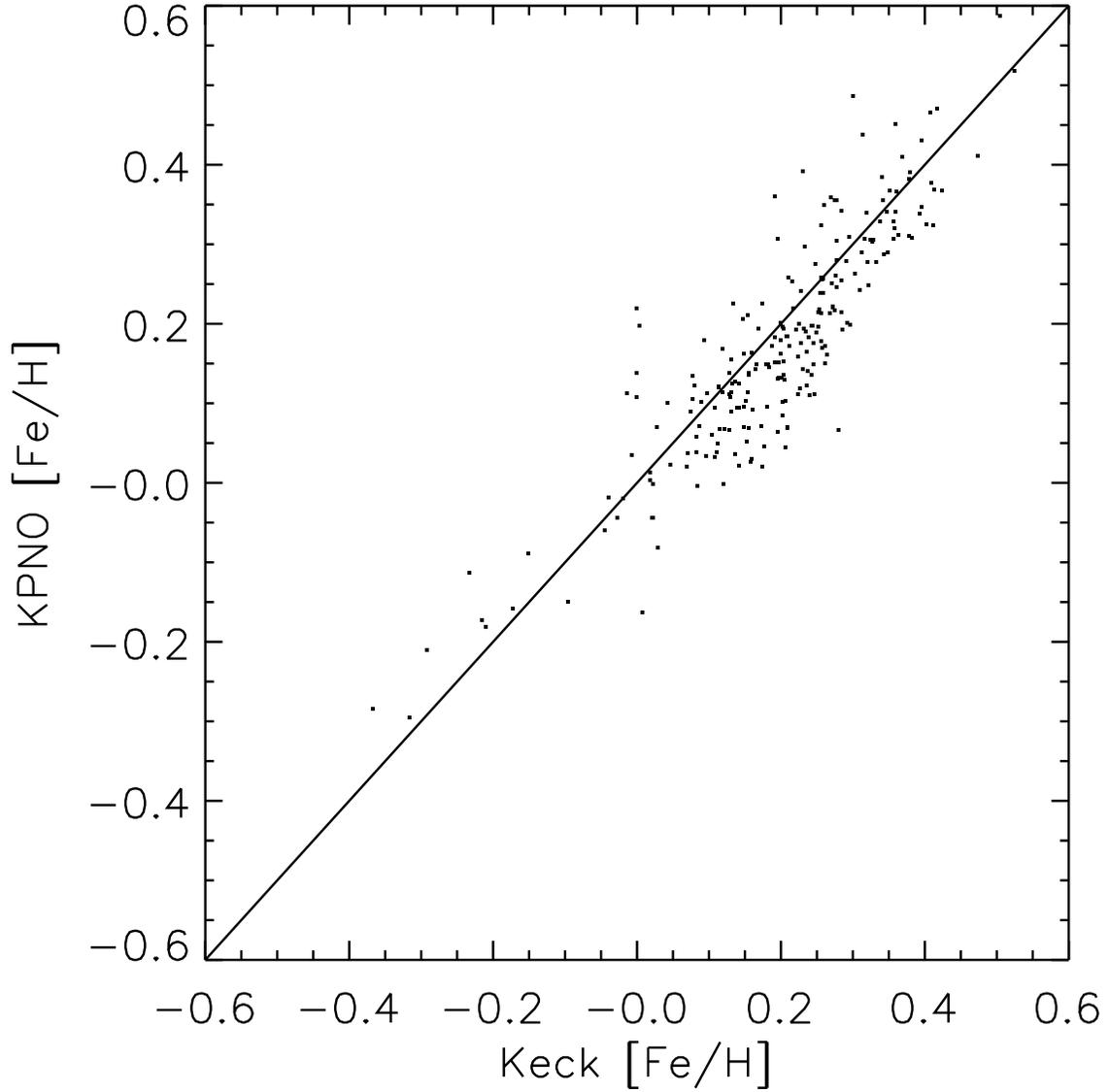}
\caption{Verification of Paper 1 [Fe/H] calibration.  For stars already
observed as part of the Keck/Subaru/Magellan planet search at the time
of this writing, [Fe/H] from the Lick Indices calibration is plotted as
a function of Keck [Fe/H].  The measurements match well, with the
standard deviation of (Lick index [Fe/H]) - (Keck [Fe/H]) at 0.07 dex.}
\label{fehkecktest}
\end{figure}

At the time of this writing, 233 of the stars observed at KPNO had
subsequently been observed with the Keck HIRES spectrograph, and [Fe/H]
measured.  In Figure \ref{fehkecktest}, KPNO [Fe/H] values are compared
with the Keck [Fe/H] measurements.  The standard deviation of $({\rm
[Fe/H]}_{\rm KPNO})$ - $({\rm [Fe/H]}_{\rm Keck})$ is 0.07 dex, as given
in Paper 1, and the center of this distribution is -0.03 dex.  As this
0.03-dex offset in the [Fe/H] zero point is robust in the range $0.00
\leq {\rm [Fe/H} \leq 0.25$, a critical range for planet searches, we
suggest measurements in future surveys using the Paper 1 method be
corrected by this value.  We also note that the training set for the
[Fe/H] calibration consisted of FGK dwarfs with approximately solar
composition: we do not expect the calibration to retain its precision or
accuracy if used on stars with a different abundance mixture.

Our test of the Paper 1 calibrations demonstrates their precision and
ease of use.  A single low-resolution spectrum, obtained with a
40-second exposure for a $V = 8$ star at the 2.1m telescope, leads to
[Fe/H] measurements that rival the precision of the high-resolution
spectroscopy in the \cite{cayreldestrobel} compilation.  (Of course,
since a high-resolution spectrum can be used to measure the abundances
of many elements, the Paper 1 calibrations certainly do not obviate the
need for high-resolution spectroscopy in characterizing stellar
populations.)  We measure Lick indices using a publicly available code
and use simple linear transformations, based on observations of stars in
the catalogs of \cite{W94} and \cite{wo97}, to place our measurements on
the published Lick system.  Of order 30 observations of Lick standards
per observing run are enough to define transformations onto the Lick
system, and a further $\sim 30$ observations of VF05 stars per observing
run verify the accuracy of the Paper 1 calibrations for each new data
set.  During the observing run of 2004 August 27-September 2 (the only
one of our observing runs where we did not lose time due to poor
weather), we were able to screen 984 potential planet-search targets, in
addition to observing 90 VF05 stars to improve the calibrations.
Although some pre-screening based on broadband photometry is necessary
to make sure targets are Population I FGK dwarfs, our calibrations make
high-throughput observing programs that return precise measurements of
stellar atmospheric parameters possible.

\section{Results: Planet-Search Targets}
\label{results}

The goal of our KPNO observing program was to identify stars that are
cool enough for successful radial-velocity measurements, and metal-rich
enough to have high probabilities of planet detection.  The ideal upper
temperature limit of planet-search targets is 6000K, because hotter
stars tend to be rapidly rotating.  Rapid rotators have broad spectral
lines that interfere with measuring precise radial velocities.  Planet
searches have been successful for stars in the range 6000-6400K,
spectral type F5-F9, although these stars can exhibit $\delta$
Scuti-type quasi-periodic velocity variations that exceed estimates of
stellar jitter (Galland et al.  2006).  For late F stars, care must be
taken to ensure that periodic radial velocity variations are maintained
for several periods, so that that stellar pulsations do not masquerade
as short-period planets.  Well-known examples of planet hosts within the
temperature range 6000-6400K are $\upsilon$ And b (HD 9826 b; Butler et
al. 1997), HD 209458 (Henry et al. 2000, Charbonneau et al. 2000) and
$\tau$ Boo (HD 120136; Butler et al. 1997).

\begin{figure}
\plotone{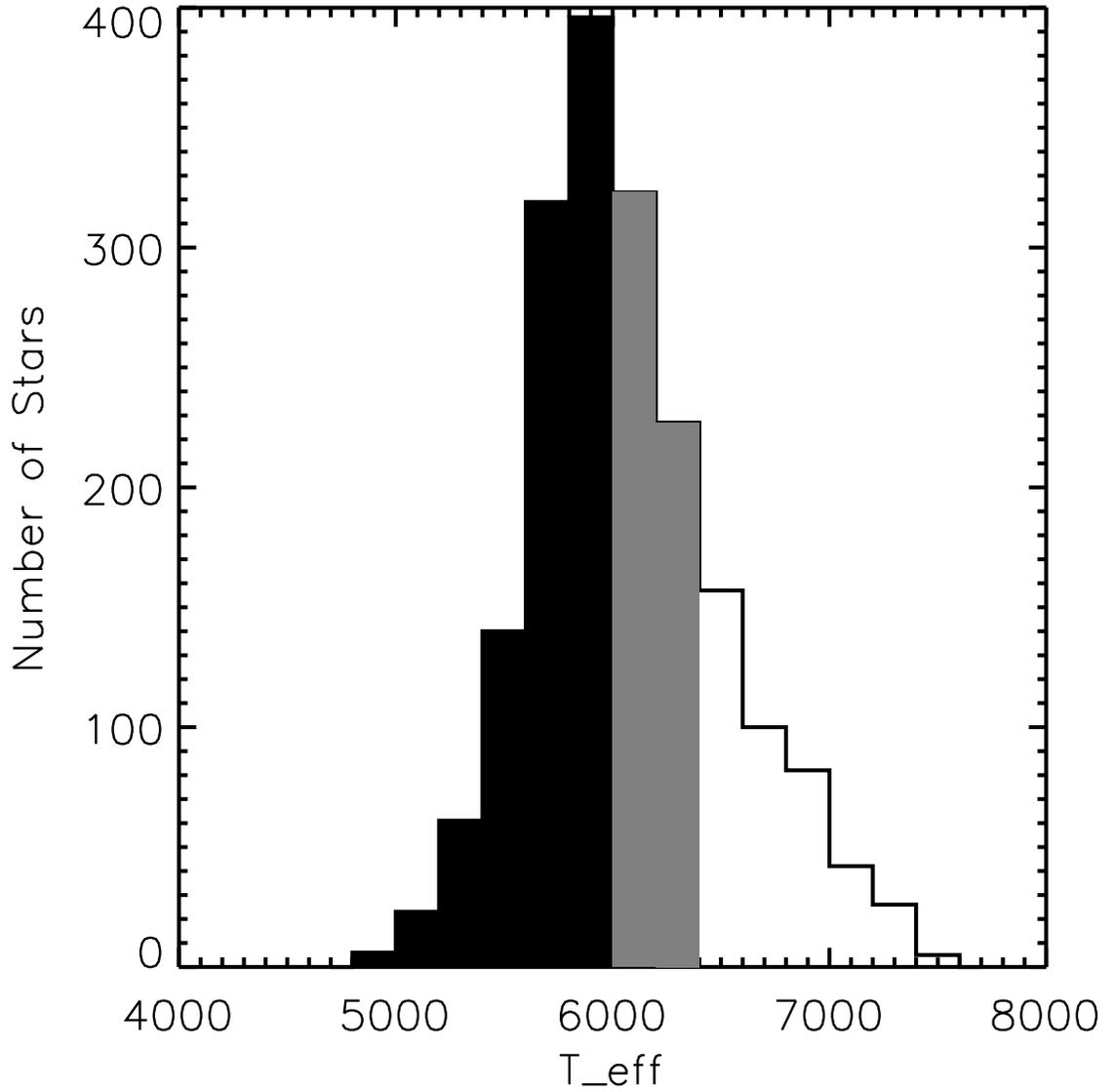}
\caption{Histogram of effective temperature of the potential
planet-search targets screened at KPNO.  Planet searches are most
effective for targets with $T_{\rm eff} \le 6000$K (shaded black).
$50\%$ of our targets fall into this category.  $78\%$ of our targets
are cooler than $T_{\rm eff} \le 6400$K (shaded gray), where planet
searches still get good results.}
\label{teffhist}
\end{figure}

Stars hotter than 6400K have weak metal lines unsuitable for
high-precision radial-velocity fits.  In Paper 1, we give the ranges in
$T_{\rm eff}$, [Fe/H] and log $g$ covered by the training sets from
which the calibrations were built: $4100 {\rm K} < T_{\rm eff} < 6400
{\rm K}$, $-0.95 \; {\rm dex} < {\rm [Fe/H]} < 0.5 \; {\rm dex}$, and
$4.0 \; {\rm dex} < \log \, g < 5.1 \; {\rm dex}$.  Since the $T_{\rm
eff}$ calibration is stable to moderate extrapolation beyond the
published range, it can reliably identify stars that are hotter than our
upper temperature limit.  946 of 1907, or $50\%$, of the stars screened
meet the ideal temperature condition of $T_{\rm eff} \le 6000$K, and
1495 of our stars, or $78\%$, are cooler than 6400K.  Figure
\ref{teffhist} shows the $T_{\rm eff}$ distribution of the N2K targets.

\begin{figure}
\plotone{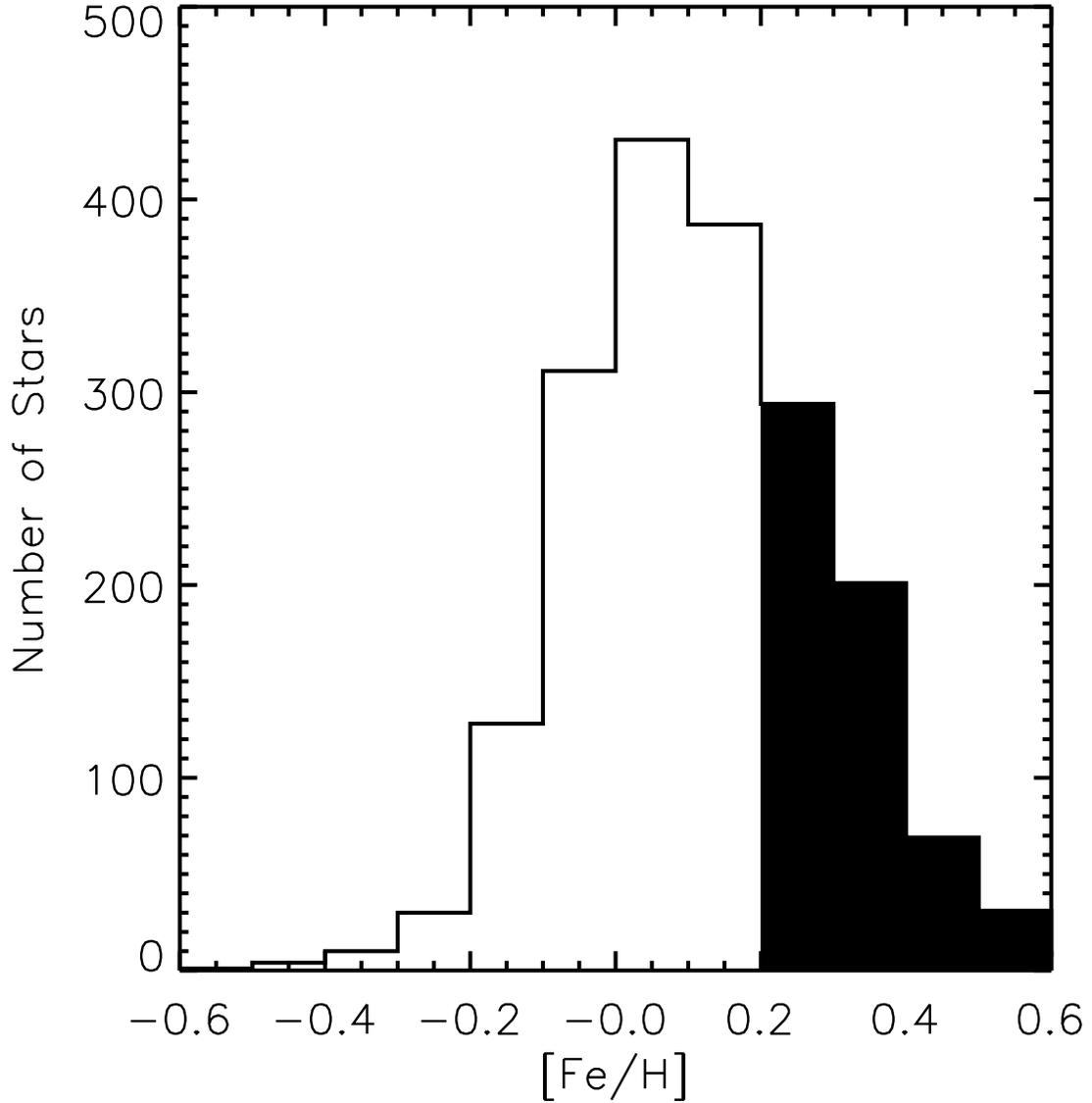}
\caption{Histogram of metallicity of the potential planet-search targets
screened at KPNO.  $32\%$ of the stars we surveyed have ${\rm [Fe/H]}
\geq 0.2$, corresponding to a $10\%$ or greater chance of harboring a
detectable planet.}
\label{fehhist}
\end{figure}

The KSM planet search is primarily focused on stars with ${\rm [Fe/H]}
\geq 0.2$ dex, which have a $10\%$ or greater probability of having a
gas giant planet (Fischer \& Valenti 2005).  605, or $32\%$ of the stars
we screened, have ${\rm [Fe/H]} \geq 0.2$ dex.  Of the 946 stars cooler
than 6000K, 284 have ${\rm [Fe/H]} \geq 0.2$ dex.  431 stars with
$T_{\rm eff} \leq 6400$K have ${\rm [Fe/H]} \geq 0.2$ dex.  Based on the
planet-metallicity correlation reported by \cite{fv05}, the 284 ideal
targets we have identified should harbor $\sim 17$ giant planets
detectable by Doppler searches, including $\sim 3$ hot Jupiters.  The
431 stars identified with $T_{\rm eff} \leq 6400$K and ${\rm [Fe/H]}
\geq 0.2$ should contain $\sim 30$ detectable planets, including 4-5
hot Jupiters.  3 planets have already been discovered among the stars
surveyed at KPNO (HIP 14810 is a double-planet system; see Wright et al.
[2007]).  Figure \ref{fehhist} shows the [Fe/H] distribution of the N2K
targets.  Table 3 contains the atmospheric parameters for
the 1907 stars observed by the N2K KPNO program.

The planet detection rate per hour of Doppler surveys is limited by the
exposure time required to reach high $S / N$.  Although increasing the
metallicity of targets by 0.1 dex increases the probability of planet
detection around each star by $58\%$ (Fischer \& Valenti 2005),
brightening targets in the $V$ band by one magnitude means 2.5 times as
many stars can be observed.  Thus, a planet search focused on stars with
$V = 8$ and ${\rm [Fe/H]} = 0.1$ dex should detect more planets than a
search allotted equal observing time, but targeting stars with ${\rm
[Fe/H]} = 0.2$ and $V = 9$.  The targets surveyed at KPNO, members of
the Hipparcos catalog (Perryman et al. 1997), were chosen from the
brightest stars available that have not already been searched for
planets.  Instead of dipping into the voluminous Tycho II catalog (H\o g
et al. 1998) to select only stars with super-Solar metallicity
estimates, we targeted Hipparcos stars with metallicity estimates from
either the N2K broadband or the \cite{nordstrom} $ubvy\beta$
calibrations of ${\rm [Fe/H} \geq 0.0$ dex.  Stars with both ${\rm
[Fe/H]}_{\rm bb} < 0.0$ dex and ${\rm [Fe/H]}_{ubvy\beta} < 0.0$ dex
(where ${\rm [Fe/H]}_{\rm bb}$ is the metallicity measured from the N2K
broadband calibration), or stars without $ubvy\beta$ photometry and with
${\rm [Fe/H]}_{\rm bb} < 0.0$ dex, were not considered for the KPNO
survey.  With this selection procedure, we could (1) enable the KSM
planet-search team to identify bright stars with metallicity slightly
below our ideal range, and (2) find stars with ${\rm [Fe/H]} \geq 0.2$
dex that were missed by the N2K broadband or $ubvy\beta$ calibrations.
(For a description of the miss and false-positive rates of the
photometric calibrations, see \S \ref{calcompare}.)


Choosing planet-search targets with astrometric distance measurements
vastly improves the ability to derive the stars' physical parameters,
such as mass and, by extension, the semimajor axis of the planetary
orbit.  It is possible to solve for stellar mass and radius only from
observed $T_{\rm eff}$ and log $g$ by assuming a typical Pop I dwarf
mass-to-light ratio, but mass and radius determinations based on direct
distance measurements are far more accurate (see Valenti \& Fischer
[2005] for the procedure for calculating mass, radius and luminosity
based on observed $T_{\rm eff}$, [M/H], log $g$ and distance).  We
targeted stars in the Hipparcos catalog for the KPNO program not only
because of their relative brightness in comparison with the more
numerous Tycho II stars, but because they have astrometric distance
measurements.  Most of the Hipparcos stars are too faint for the
forthcoming 2.4m Automated Planet Finder telescope, which will seek
low-mass companions around stars that have already been observed by
Doppler surveys.  By searching the Hipparcos catalog for giant planets,
the KSM planet search fills an important niche: stars that have not yet
been surveyed for planets, are too faint for small telescopes, and have
astrometric distance measurements.

One final concern for planet searches is whether the target stars are
members of multiple systems, because precise measurement of radial
velocities is impeded when two spectra enter the slit. From a
theoretical perspective, the orbit of a giant planet in a binary system
can only be stable if it is circumbinary with a semimajor axis $a_{\rm
pl}$ more than $\sim 3$ times the mean stellar separation $a_*$, or
around one star only with $a_{\rm pl} \lesssim (1/3) a_*$.  Multiple
star systems may therefore have protoplanetary disks that are unstable
for giant planet formation.  An unusual case is the hot Jupiter orbiting
the primary of the triple system HD 188753 (Konacki 2005).  In Table
3, we note binary systems present in the SIMBAD astronomical
database\footnote{\copyright ULP/CNRS - Centre de Donn\'{e}es
astronomiques de Strasbourg}.  We also note stars that appeared to have
a close companion on the 2.1m telescope slit camera, with a field of
view $5 \arcsec \times 1 \farcs 3$, the slit width.




\section{Performance of Broadband and $ubvy\beta$ Calibrations}
\label{calcompare}

In this section, we assess the N2K observing strategy.  In brief, this
consists of starting with \cite{mark} broadband (hereafter N2K
broadband) or \cite{nordstrom} $ubvy\beta$ (hereafter $ubvy\beta$)
[Fe/H] estimates, refining these measurements with low-resolution
spectroscopy at KPNO, placing the brightest and most metal-rich stars
from the KPNO survey on the KSM planet-search target list, and finally,
obtaining photometric observations to check hot-Jupiter candidates for
transits.  We report the numbers of misses and false positives produced
by the photometric [Fe/H] and $T_{\rm eff}$ calibrations.  (\cite{mark}
did not build a photometric log $g$ calibration.)  Finally, we compare
the precision of the N2K broadband, $ubvy\beta$, and Paper 1
calibrations, and discuss the benefits of obtaining the extra precision
offered by low-resolution spectroscopy before proceeding to planet
searches.

Although the broadband [Fe/H] calibration has uncertainty $\sigma \leq
0.17$ dex for stars brighter than $V = 9$, beyond this magnitude limit
photometric errors increase the uncertainty of [Fe/H] estimates to
$\sigma \geq 0.3$ dex.  A target with $V = 9$ and ${\rm [Fe/H]}_{\rm bb}
= 0.2$, or $P_{\rm planet} = 0.08$, has a $16\%$ chance of having a true
metallicity ${\rm [Fe/H]} \leq -0.1$ dex and $P_{\rm planet} = 0.02$, a
$75\%$ reduction in the probability of planet detection.  This outcome
is a ``false positive,'' where a star appears to be above our ideal
[Fe/H] for planet-search targets, but in fact is not.  Since we are also
looking for cool stars, another type of false positive is when a star is
identified as having $T_{\rm eff} < 6400$K, but is in fact hotter.  When
targets are bright stars with astrometric distance measurements---ideal
planet-search targets in every way except possibly metallicity, which is
unknown---another problematic outcome is a ``miss,'' where a star with
that is truly metal-rich is identified as metal-poor.  A miss also
results when a cool star is mistakenly identified as having $T_{\rm eff}
\geq 6400$K.  Although the KSM planet search does observe stars
with ${\rm [Fe/H]} \le 0.2$, its main focus is stars with super-Solar
metallicity, ${\rm [Fe/H]} \geq 0.2$, so we will make this our
metallicity cutoff.  We set our temperature cutoff at $T_{\rm eff} =
6400$K, the point at which metal lines become too weak for planet
searches.

\begin{figure}
\plotone{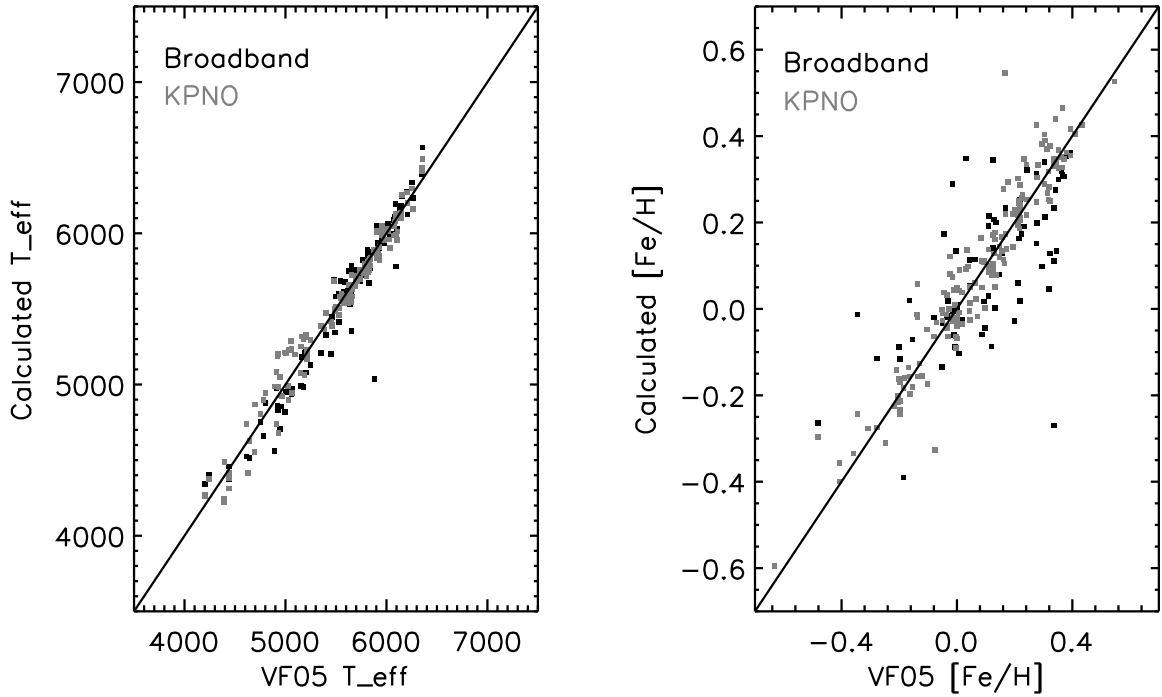}
\caption{Performance of N2K broadband and Paper 1 calibrations for
stars in common with VF05.  Left: $T_{\rm eff}$; Right: [Fe/H].  Black
symbols correspond to output of broadband calibration, and gray symbols
indicate results of Lick indices calibration.  The solid line in each
plot shows a 1:1 correlation.}
\label{bbtestset}
\end{figure}

``Hits,'' the most desirable outcomes of pre-planet-search screening,
are stars with both ${\rm [Fe/H]}_{\rm bb}$ and ${\rm [Fe/H]}_{\rm KPNO}
\geq 0.2$ (where ${\rm [Fe/H]}_{\rm KPNO}$ is metallicity measured from
using the Paper 1 calibration on the KPNO spectra).  A hit for the
temperature calibration results when both $T_{\rm eff,bb}$ and $T_{\rm
eff,KPNO}$ are lower than 6400K.  The N2K broadband [Fe/H] calibration
produced 485 hits, while the broadband $T_{\rm eff}$ calibration had
1388 hits.  False positives are the least desirable outcome because they
may lead to wasted time at large telescopes if stars are not screened
with the Paper 1 calibrations.  According to our KPNO $T_{\rm eff}$
measurements, the N2K broadband $T_{\rm eff}$ calibration only had 46
false positives, just over $3\%$ of the stars it identified as cooler
than our 6400K temperature limit.  However, the broadband [Fe/H]
calibration produced 337 false positives.  If stars were selected for
the KSM planet search based on broadband metallicity measurements alone,
an unacceptable $38\%$ of the stars on the target list would be false
positives, as opposed to only $9\%$ for the stars selected from KPNO
observations plus broadband screening.  (The Paper 1 calibrations would
have much higher false positive and miss rates if tested on stars that
had not been subject to photometric screening, since they are only valid
for FGK dwarfs of approximately solar abundance mixture.)  Figure
\ref{bbtestset} compares the performance of the broadband and Lick index
calibrations for our set of VF05 stars observed at KPNO.  Although both
$T_{\rm eff}$ calibrations perform about equally well, the [Fe/H]
measurements from KPNO spectra are noticeably more precise than those
from the broadband [Fe/H] calibration.

\begin{figure}
\plotone{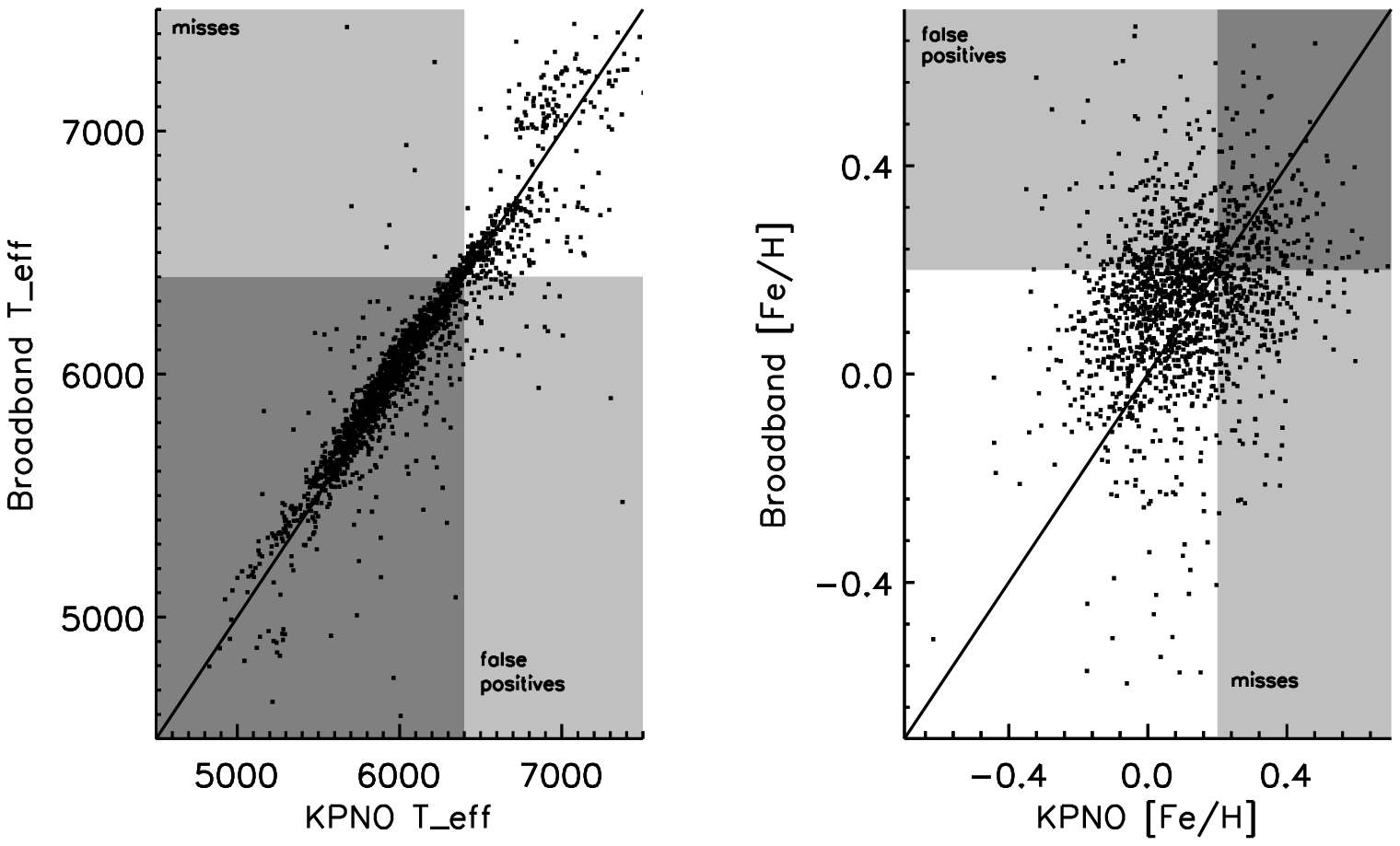}
\caption{$T_{\rm eff}$ and [Fe/H] from broadband calibration as a
function of values measured from the Paper 1 calibrations.  Although the
$T_{\rm eff}$ measurements from KPNO spectra give little gain in
precision over the \cite{mark} values, [Fe/H] measurements from KPNO
give a gain in precision of between 0.08 and 0.23 dex.  The Paper 1
[Fe/H] calibration was able to reject 52 stars identified by the
broadband calibration as being extremely metal-rich, but which in fact
have $84\%$ or higher probability of having ${\rm [Fe/H]} < 0.2$.}
\label{bbprogram}
\end{figure}

Of course, not all false positives truly lead to wasted time on large
telescopes: a star with [Fe/H] = 0.18 dex measured by the Paper 1
calibration, but 0.22 dex as measured by the N2K broadband calibration,
is still a desirable planet-search target.  We most want to flag false
positives with dramatic metallicity overestimates, $\sim 0.2$ dex or
more.  Also, occasionally the broadband [Fe/H] measurement will be
closer to the true value than the KPNO measurement: for the example
quoted above, the 0.04 dex metallicity difference between the two
estimates is within the error of the Paper 1 calibration.  To get an
idea of how often the broadband [Fe/H] calibration produces true
misidentifications, we count the number of stars with ${\rm [Fe/H]}_{\rm
bb} \geq 0.2 + \sigma_{\rm bb}$ but ${\rm [Fe/H]}_{\rm KPNO} < 0.2 -
\sigma_{\rm KPNO}$. These are the false positives for which the error
bars of the two calibrations do not overlap.  There are 34 of these true
false positives among the stars surveyed at KPNO; the KSM planet-search
target list is therefore much cleaner as a result of having been vetted
by the KPNO observations.  Figure \ref{bbprogram} shows a comparison of
N2K broadband and KPNO [Fe/H] and $T_{\rm eff}$ measurements for all the
stars screened at KPNO.  Here again, we see that the temperature
measurements from both calibrations match well, and the real gain
provided by the KPNO observations is in precision of [Fe/H]
measurements.

To measure [Fe/H] from $ubvy\beta$ photometry, \cite{nordstrom} used the
$ubvy\beta$-metallicity calibration of \cite{sn89} (hereafter SN89),
which has precision $\sigma = 0.13$ dex.  For very red G and K dwarfs,
however, the SN89 calibration produces large systematic errors in
metallicity (Twarog, Anthony-Twarog \& Tanner 2002); \cite{nordstrom}
thus derive a new $ubvy\beta$-metallicity calibration for cool G and K
dwarfs.  The $ubvy\beta$ [Fe/H] values were checked against the
spectroscopic metallicities of \cite{taylor03}, \cite{edvardsson93} and
\cite{chen00} and found to be in good agreement with each the values in
each catalog.

\begin{figure}
\plotone{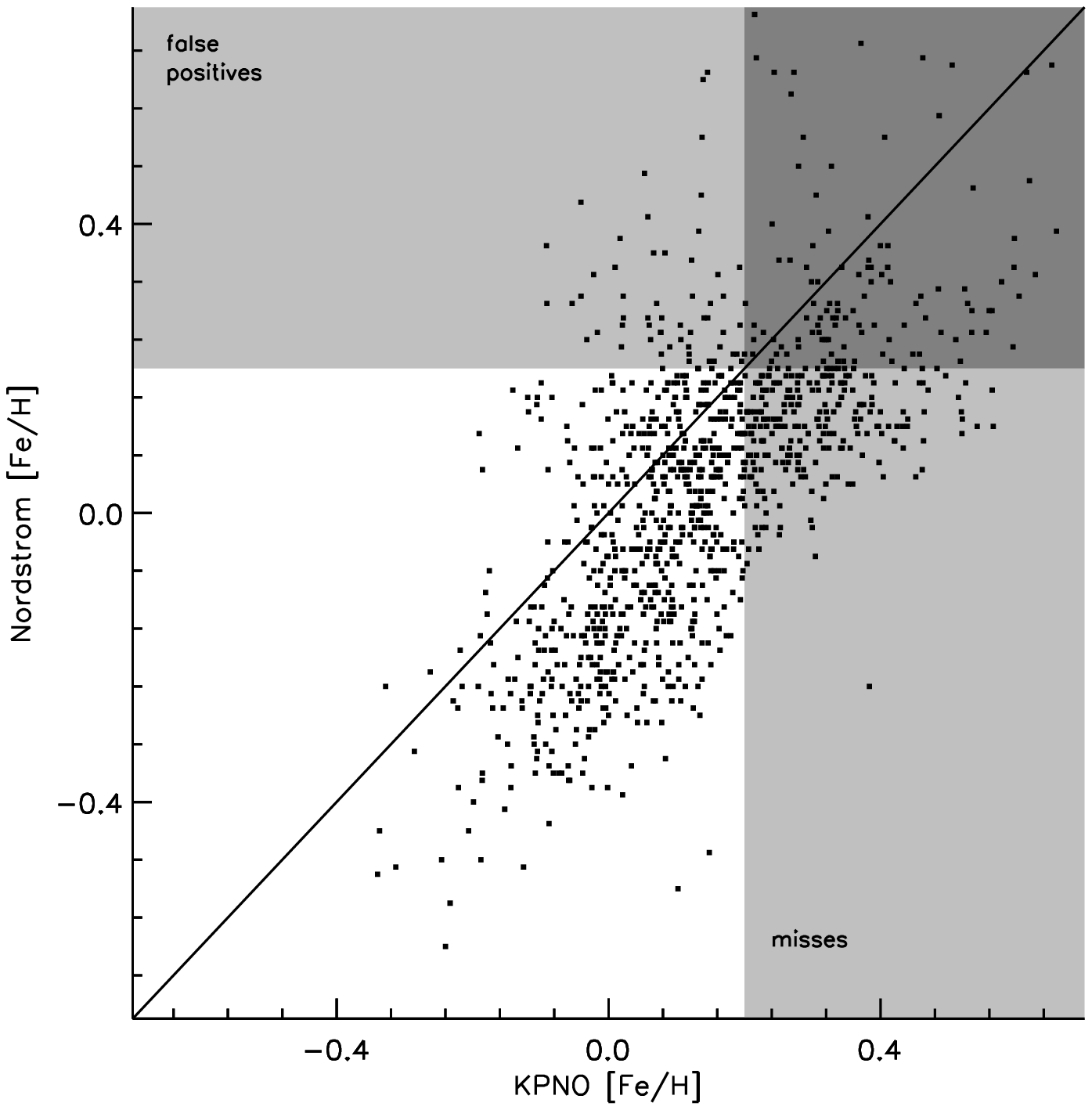}
\caption{[Fe/H] from \cite{nordstrom} $ubvy\beta$ photometry as function
of values measured from Lick indices and Paper 1 calibration.  The
$ubvy\beta$ metallicities were mostly calculated with the \cite{sn89}
calibration, which was built for Pop II stars and is known to
underestimate [Fe/H] for metal-rich stars.  The scatter above [Fe/H] =
0.2 and the systematic [Fe/H] underestimates show that the Paper 1
calibrations provide a substantial improvement in precision and accuracy
over the $ubvy\beta$  metallicities.}
\label{nordcompare}
\end{figure}

We have two reasons for following up the fainter $ubvy\beta$ [Fe/H]
estimates with low-resolution spectroscopy, and not adding the most
metal-rich stars directly to the KSM planet search: (1) for stars with
both N2K broadband and $ubvy\beta$ [Fe/H] estimates, there was often a
discrepancy between the two calibrations of 0.2 dex or more, and (2)
very few metal-rich stars were used to build the SN89 calibration.
\cite{sarah} noted [Fe/H] underestimates as a problem for metal-rich
stars in SN89: a residual histogram for stars with ${\rm [Fe/H]}_{\rm
spec} \geq 0.0$ is centered at -0.08 dex.  This systematic underestimate
would lead to many misses.  Of the 1907 stars surveyed at KPNO, 1052
have $ubvy\beta$ metallicity estimates and 184 stars have both
$ubvy\beta$ and N2K broadband metallicity estimates.  We count 227
misses and 61 false positives among the $ubvy\beta$ stars we surveyed,
for a miss rate of $22\%$.  22 of the misses and 24 of the false
positives were cases where the error bars of the $ubvy\beta$ and Paper 1
calibrations did not overlap.  Figure \ref{nordcompare} shows a
comparison of the $ubvy\beta$ and KPNO [Fe/H] measurements.

Although the Paper 1 [Fe/H] calibration has quite high
precision---enough to justify creating an observing program to screen
planet-search targets---it has a limited range of use, only $-0.95 \leq
{\rm [Fe/H]} \leq 0.5$ dex.  Some type of photometric metallicity
calibration is therefore absolutely necessary to weed out Population II
stars and ensure that targets for low-resolution spectroscopy fall in
the appropriate metallicity range.  The \cite{mark} and \cite{nordstrom}
calibrations both perform this task admirably.  As a result of the KPNO
survey, we know that the N2K broadband $T_{\rm eff}$ calibration gives
precise measurements even for stars dimmer than the published magnitude
limit of $V = 9$, where photometric error is comparable to the internal
calibration error.  For future low-resolution spectroscopic surveys of
this type, we can rely on the N2K broadband calibration to reject hot
stars without any further verification.  The N2K strategy of beginning
with photometric [Fe/H] measurements, refining them by measuring Lick
indices and using the Paper 1 calibrations, and finally placing the
cool, metal-rich stars in the KSM planet search has so
far been profitable, leading to the discovery of 3 new planets.  We
expect the N2K target list to be yielding new planet discoveries for
some time.

\section{Conclusion}
\label{conclusion}

We have calculated high-precision atmospheric parameters for 1907 FGK
dwarfs in the Solar neighborhood.  The ideal planet-search targets we
identified will feed the Keck, Subaru and Magellan planet searches for
the next 2 years.  The 284 best targets, those with ${\rm [Fe/H]} \geq
0.2$ dex (for a $\geq 10\%$ probability of harboring a detectable
planet) and $T_{\rm eff} \leq 6000$K, should yield $\sim 17$ new planet
discoveries.  The entire catalog of 1907 stars should eventually lead to
$> 60$ planet discoveries, including 8-9 hot Jupiters.  Two hot Jupiters
have already been found among our 1907 survey targets.  As 10 of 48
known short-period planets display detectable transits, we hope that 1
or 2 additional transits may be found among the stars surveyed for this
work.  The high-quality planet-search targets identified by our
low-resolution spectroscopic survey will keep the planet detection rate
of the Keck/Subaru/Magellan program high, even as it pushes out to
larger distances and fainter stars.

The N2K pipeline is an efficient and successful way to identify stars
likely to host detectable planets.  With the information from a single
low-resolution spectrum, the calibrations in Paper 1 can provide
atmospheric parameter measurements for any Pop I dwarf with precision
rivaling some high-resolution surveys.  Indeed, only $9\%$ of the stars
identified as having ${\rm [Fe/H]}_{\rm KPNO} \geq 0.2$, and observed
once at Keck at the time of this writing, were found to have ${\rm
[Fe/H]}_{\rm Keck} < 0.2$.  The \cite{mark} temperature calibration is
highly precise---$\sigma \leq 85$K for stars $V < 10$---and can be used
on any dwarf star with $BVJHK$ photometry, which includes more than
100,000 stars in the Tycho II catalog (H\o g et al. 1998).  The N2K
broadband and \cite{nordstrom} $ubvy\beta$ metallicity calibrations are
provide excellent first estimates of [Fe/H] and enable us to reject
low-metallicity targets from our second-tier screening with
low-resolution spectroscopy.

The main source of uncertainty in the N2K broadband [Fe/H] calibration
is simply photometric error.  A photometric catalog in which every
measurement has the same $S / N$ might make screening planet-search
targets with low-resolution spectroscopy unnecessary.  Indeed, the
$ubvy$-metallicity calibration of \cite{sarah}, with precision $\sigma =
0.10$ dex, was successful at identifying the first generation of Keck
planet-search targets.  Once the Hipparcos catalog has been thoroughly
searched for planets, empirical metallicity calibrations could be
created using $ugriz$ photometry from the Sloan Digital Sky Survey
(Adelman-McCarthy et al. 2006).  With over 6670 deg$^2$ surveyed,
metal-rich, nearby stars from the SDSS catalog could feed automated
planet-searches for generations to come, and enable such ambitious
programs as taking the planet census of the entire Solar neighborhood.

{\it Facility:} \facility{KPNO:2.1m}


\clearpage
\begin{deluxetable}{lrrrc}
\tabletypesize{\tiny}
\tablecaption{Matching the Lick system: Linear transformations from
observed to published Lick indices and index errors
\label{licktrans} }
\tablehead{
\colhead{Index} & \colhead{Slope} & \colhead{Intercept} &
\colhead{Error} & \colhead{\# Rejected\tablenotemark{a}} }
\startdata
Ca4227 & 1.101 & -0.309 & 0.210 & 1 \\
 & 0.844 & 0.108 & 0.274 & 1 \\
 & 1.072 & -0.103 & 0.194 & 0 \\
G4300 & 1.229 & -0.911 & 0.372 & 3 \\
 & 1.410 & -1.723 & 0.296 & 1 \\
 & 1.782 & -3.276 & 0.481 & 1 \\
H$\gamma_F$ & 1.068 & -0.019 & 0.520 & 3 \\
 & 1.199 & -0.158 & 0.446 & 1 \\
 & 0.980 & 0.128 & 0.230 & 1 \\
Fe4383 & 0.975 & -0.575 & 0.693 & 1 \\
 & 0.928 & 0.300 & 0.615 & 0 \\
 & 1.113 & -0.254 & 0.543 & 2 \\
Fe4531 & 0.988 & -0.401 & 0.402 & 1 \\
 & 0.948 & -0.184 & 0.362 & 0 \\
 & 1.097 & -0.568 & 0.287 & 1 \\
Fe4668 & 1.067 & -0.234 & 0.607 & 2 \\
 & 1.106 & -0.533 & 0.554 & 1 \\
 & 1.082 & -0.249 & 0.529 & 2 \\
H$\beta$ & 0.991 & -0.133 & 0.161 & 2 \\
 & 0.958 & -0.045 & 0.359 & 1 \\
 & 0.975 & -0.128 & 0.198 & 2 \\
Fe5015 & 1.055 & -0.278 & 0.501 & 0 \\
 & 0.962 & -0.010 & 0.632 & 0 \\
 & 1.087 & -0.465 & 0.546 & 0 \\
Mg$_2$ & 1.036 & 0.033 & 0.010 & 1 \\
 & 0.995 & 0.047 & 0.009 & 0 \\
 & 1.021 & 0.040 & 0.010 & 0 \\
Mg {\it b} & 1.376 & 0.518 & 0.355 & 2 \\
 & 1.344 & 0.499 & 0.354 & 2 \\
 & 1.231 & 0.606 & 0.457 & 0 \\
Fe5270 & 1.186 & -0.257 & 0.308 & 0 \\
 & 1.044 & -0.072 & 0.261 & 0 \\
 & 1.107 & -0.113 & 0.220 & 0 \\
Na5895 & 1.149 & -0.279 & 0.211 & 2 \\
 & 0.933 & 0.354 & 0.211 & 1 \\
 & 1.135 & -0.143 & 0.282 & 0 \\
\enddata
\tablenotetext{a}{Number of points rejected from final compuatation of
transformation}
\tablenotetext{b}{Top row gives transformations for data taken in August
2004; middle row gives transformations for data taken in March 2005;
bottom row gives transformations for data taken in April 2005}
\end{deluxetable}



\begin{thebibliography}

\bibitem[Adelman-McCarthy et al.(2006)]{sloan} Adelman-McCarthy, J. K.,
Ag\"{u}eros, M. A., Allam, S. S., et al. 2006, \apjs, 162, 38

\bibitem[Ammons et al.(2006)]{mark} Ammons, S. M., Robinson, S. E.,
Strader, J., Laughlin, G., Fischer, D., \& Wolf, A. 2006, \apj, 638,
1004

\bibitem[Butler et al.(1997)]{tauboo} Butler, R. P., Marcy, Geoffrey W.,
Williams, E., Hauser, H., \& Shirts, P. 1997, \apj, 474, L115

\bibitem[Cardiel et al.(1998)]{cardiel98} Cardiel, N., Gorgas, J.,
Cenarro, J., \& Gonz\'{a}lez, J. Jes\'{u}s. 1998, \aaps, 127, 597

\bibitem[Cayrel de Strobel, Soubiran \& Ralite(2001)]{cayreldestrobel}
Cayrel de Strobel, G., Soubiran, C., \& Ralite, N. 2001, \aap, 373, 159

\bibitem[Charbonneau et al.(2000)]{charbonneau00} Charbonneau, D.,
Brown, T. M., Latham, D. W., \& Mayor, M. 2000, \apj, 529, L45

\bibitem[Chen et al.(2000)]{chen00} Chen, Y. Q., Nissen, P. E., Zhao,
G., Zhang, H. W., \& Benoni, T. 2000, \aaps, 141, 491

\bibitem[Edvardsson et al.(1993)]{edvardsson93} Edvardsson, B.,
Andersen, J., Gustafsson, B., Lambert, D. L., Nissen, P. E., \& Tomkin,
J. 2003, \aaps, 102, 603

\bibitem[Fischer et al.(2005)]{fischer05} Fischer, D. A., Laughlin, G.,
Butler, P., Marcy, G., Johnson, J., Henry, G., Valenti, J., Vogt, S.,
Ammons, M., Robinson, S., Strader, J., et al. 2005, \apj, 620, 481

\bibitem[Fischer et al.(2006)]{fischer06} Fischer, D. A., Laughlin, G.,
Marcy, G. W., Butler, R. P., Vogt, S. S., Johnson, J. A., Henry, G. W.,
McCarthy, C., Ammons, S. M., Robinson, S., Strader, J., et al. 2006,
\apj, 637, 1094

\bibitem[Fischer \& Valenti(2005)]{fv05} Fischer, D. A., \& Valenti, J.
2005, \apj, 622, 1102

\bibitem[Fortney et al.(2005)]{fortney05} Fortney, J. J., Saumon, D.,
Marley, M. S., Lodders, K., \& Freedman, R. 2005, AAS/Division for
Planetary Sciences Meeting Abstracts, 37,

\bibitem[Galland et al.(2006)]{galland06} Galland, F., Lagrange, A.-M.,
Udry, S., Chelli, A., Pepe, F., Beuzit, J.-L., \& Mayor, M. 2006, \aap,
447, 355


\bibitem[Henry et al.(2000)]{henry00} Henry, G. W., Marcy, G. W.,
Butler, R. P., \& Vogt, S. S. 2000, \apj, 529, L41

\bibitem[H\o g et al.(1998)]{tycho} H\o g, E., Kuzmin, A., Bastian, U.,
Fabricius, C., Kuimov, K., Lindegren, L., Makarov, V. V., \& Roeser, S.
1998, \aap, 335, L65

\bibitem[Hubickyj, Bodenheimer \& Lissauer(2005)]{hubickyj05} Hubickyj,
O., Bodenheimer, P., \& Lissauer, J. J. 2005, \icarus, 179, 415


\bibitem[Konacki(2005)]{triplestar} Konacki, M. 2005, \nat, 436, 230



\bibitem[Marcy et al.(2005)]{marcy05} Marcy, G., Butler, R. P., Fischer,
D., Vogt, S., Wright, J. T., Tinney, C. G., \& Jones, H. R. A. 2005,
Progress of Theoretical Physics Supplement, 158, 24

\bibitem[Martell \& Laughlin(2002)]{sarah} Martell, S., \& Laughlin,
G. 2002, \apj, 577, L45

\bibitem[Nakajima et al.(2005)]{nakajima05} Nakajima, T., Morino, J.-I.,
Tsuji, T., et al. 2005, Astronomische Nachrichten, 326, 952

\bibitem[Nordstr\"{o}m et al.(2004)]{nordstrom} Nordstr\"{o}m, B.,
Mayor, M., Andersen, J., Holmberg, J., Pont, F., J\o rgensen, B.
R., Olsen, E. H., Udry, S., \& Mowlavi, M. 2004, \aap, 418, 989

\bibitem[Perryman et al.(1997)]{hipparcos} Perryman, M. A. C.,
Lindegren, L., Kovalevsky, J., Hoeg, E., Bastian, U., Bernacca, P. L.,
Cr\'{e}z\'{e}, M., Donati, F., Grenon, M., van Leeuwen, F., van der
Marel, H., Mignard, F., Murray, C. A., Le Poole, R. S., Schrijver, J.,
Turon, C., Arenou, F., Froeschl\'{e}, M., \& Petersen, C. S. 1997, \aap,
323, L49

\bibitem[Robinson et al.(2006)]{me} Robinson, S. E., Strader, J.,
Ammons, S. M., Laughlin, G., \& Fischer, D. 2006, \apj, 637, 1102 (Paper
1)


\bibitem[Sato et al.(2005)]{sato05} Sato, B., Fischer, D. A., Henry, G.
W., Laughlin, G., Butler, R. P., Marcy, G. W., Vogt, S. S., Bodenheimer,
P., Ida, S., Toyota, E., et al. 2005, \apj, 633, 465

\bibitem[Saumon \& Guillot(2004)]{sg05} Saumon, D., \& Guillot, T. 2004,
\apj, 609, 1170

\bibitem[Schuster \& Nissen(1989)]{sn89} Schuster, W. J., \& Nissen, P.
E. 1989, \aap, 221, 65

\bibitem[Skrutskie et al.(2006)]{skrutskie06} Skrutskie, M. F., Cutri,
R. M., Stiening, R., Weinberg, M. D., et al. 2006, \aj, 131, 1163

\bibitem[Taylor(2003)]{taylor03} Taylor, B. J. 2003, \aap, 398, 731

\bibitem[Trager et al.(1998)]{trager98} Trager, S. C., Worthey, G.,
Faber, S. M., Burstein, D., \& Gonz\'{a}lez, J. J. 1998, \apjs, 116, 1

\bibitem[Twarog, Anthony-Twarog \& Tanner(2002)]{twarog02} Twarog, B.
A., Anthony-Twarog, B. J., \& Tanner, D. 2002, \aj, 123, 2715

\bibitem[Valenti \& Fischer(2005)]{vf05} Valenti, J. A., \& Fischer, D.
A. 2005, \apjs, 159, 141 (VF05)

\bibitem[Williams et al.(2006)]{williams06} Williams, P. K. G.,
Charbonneau, D., Cooper, C. S., Showman, A. P., \& Fortney, J. J. 2006,
\apj, 649, 1020

\bibitem[Wright et al.(2007)]{wright07} Wright, J. T., Marcy, G. W.,
Fischer, D. A., Butler, R. P., Vogt, S. S., Tinney, C. G., Jones, H. R.
A., Carter, B. D., Johnson, J. A., McCarthy, C., \& Apps, K. 2007, \apj,
in press

\bibitem[Worthey et al.(1994)]{W94} Worthey, G., Faber, S. M.,
Gonz\'{a}lez, J. Jes\'{u}s, \& Burstein, D. 1994, \apjs, 94, 687

\bibitem[Worthey \& Ottaviani(1997)]{wo97} Worthey, G., \& Ottaviani, D.
L. 1997, \apjs, 111, 377

\end{thebibliography}
\end{document}